\documentclass[prl,twocolumn,showpacs,floatfix]{revtex4}
\usepackage[T1]{fontenc} 
\usepackage[cp1250]{inputenc}
\usepackage{graphicx} 
\usepackage[english]{babel} 
\usepackage{psfrag}
\usepackage{color} 
\newcommand{\bq}{\begin{equation}} 
\newcommand{\eq}{\end{equation}}
\newcommand{\ba}{\begin{eqnarray}} 
\newcommand{\ea}{\end{eqnarray}}
\begin{document}
\title{Explosive condensation in a mass transport model}
\author{Bart{\l}omiej Waclaw}
\author{Martin R. Evans}
\affiliation{SUPA, School of Physics and Astronomy, University of Edinburgh, Mayfield Road, Edinburgh EH9 3JZ, United Kingdom}
\noindent
\begin{abstract}
We study a far-from-equilibrium system of interacting particles, hopping between sites of a 1d lattice with a rate which increases with the number of particles at interacting sites. We find that clusters of particles, which initially spontaneously form in the system, begin to move at increasing speed as they gain particles. Ultimately, they produce a  moving
condensate which comprises a finite fraction of the mass in the system. We show that, in contrast with previously studied models
of condensation, the relaxation time to steady state decreases as an inverse power of $\ln L$ with system size $L$ and that condensation is instantenous for $L\to\infty$. 
\end{abstract}
\maketitle 

Recent studies in non-equilibrium statistical physics show that diverse phenomena such as
jamming in traffic flow \cite{chowdhury},
polydisperse hard spheres \cite{EMPT10}, wealth condensation in
macroeconomies \cite{BJJKNPZ02}, hub formation in complex networks \cite{redner}, pathological phases in quantum gravity \cite{bbpt}, and general problems of phase separation \cite{kafri} can be understood by the condensation transition. 
Condensation occurs when the global density of a conserved
quantity (mass, wealth etc.) exceeds a critical value, and manifests itself as a finite fraction of the total system mass localized in space.
A well-studied, fundamental  model  is the Zero-range Process (ZRP)
which may serve as either  a microscopic or effective description of non-equilibrium condensation
\cite{EH05,kafri,gss,godreche+luck}. In this model
particles hop to the right on a closed chain of $L$ sites with rates $u(m) \cong 1+ \gamma/m$
depending only on the number of particles $m >0$ at the departure site. The condensate, which exists in this model for
$\gamma>2$ when density of particles is above some critical value, remains static once it has formed, melting and reforming very rarely \cite{godreche+luck}. This is caused by attractive interactions between particles
expressed in $u(m)$: the more particles are in the condensate, the slower it evolves.

In this work we demonstrate a novel mechanism of non-equilibrium condensation motivated by processes such as gravitational clustering \cite{silk-white}, formation of droplets in clouds or on inclined surfaces due to collisions \cite{falkovich}, and differential sedimentation \cite{horvai}, where aggregation of particles speeds up in time as a result of increasing exchange rate of particles between growing clusters. 
For example, raindrops falling through the mist increase their velocity when gaining mass, which causes them to accrete mass even faster. 
To better understand the difference between the dynamical nature of the condensate in such processes and 
the static condensation which has previously been studied \cite{chowdhury,EMPT10,BJJKNPZ02,redner,bbpt,evans1}, we consider a microscopic model of particles hopping between sites of a 1d lattice with rate $u(m,n)\sim (mn)^\gamma$ which increases with the numbers $m,n$ of particles at interacting sites. We shall show that for $\gamma>2$
condensation occurs through a contrasting dynamical mechanism to
that previously considered --- the formation of the condensate happens
on a very fast time scale and we term it {\it explosive}.  By considering
the microscopic processes of the dynamics we show that the condensate
moves with speed $v \sim L ^\gamma$ which increases with system size $L$. 
We argue that each cluster of particles has a chance to develop into the condensate in finite time. It then follows from
extreme value statistics  that the time to form of the condensate {\em decreases} with system size as $(\ln
L)^{1-\gamma}$, in contrast to ZRP. This counter-intuitive result means that condensation is instantenous for $L\to\infty$.

\begin{figure}%
	\psfrag{m}{$m$} \psfrag{n}{$n$} \psfrag{umn}{$u(m,n)\sim (mn)^\gamma$}
	\includegraphics*[width=6.0cm]{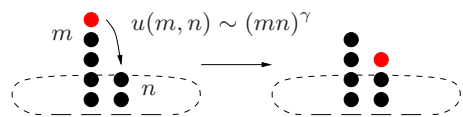}
	\caption{Model definition: a particle hops from site with $m$ to site with $n$ particles with rate $u(m,n)$.
	\label{model_def}}%
\end{figure}

\begin{figure}
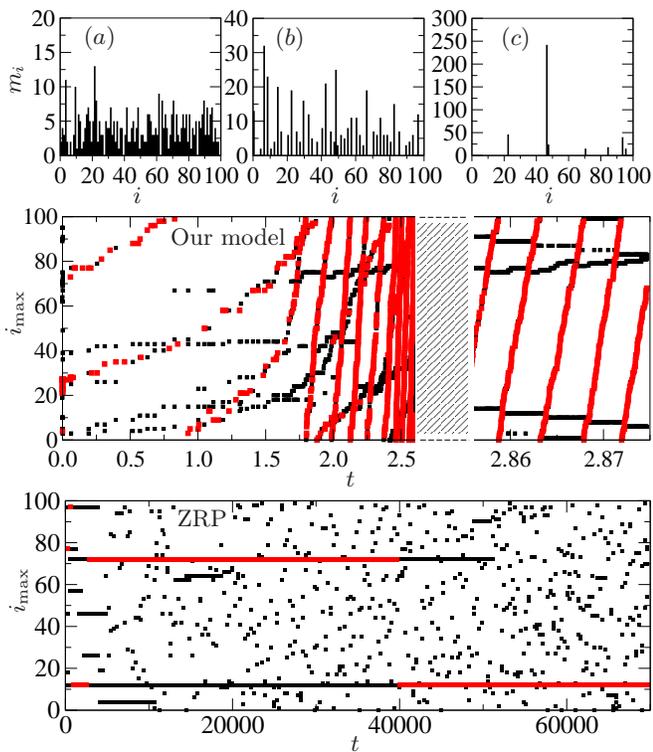
%
	\psfrag{xx}{$i$} \psfrag{yy}{$m_i$}
	\psfrag{(a)}{$(a)$} \psfrag{(b)}{$(b)$} \psfrag{(c)}{$(c)$}
	\includegraphics*[width=8.5cm]{dynamics_expl_snapshots.eps}
	\psfrag{yy}{$i_{\rm max}$}\psfrag{xx}{$t$}
	\psfrag{y1}{$i_{\rm max}$}\psfrag{x1}{$t$}
	\psfrag{our_model}{Our model}
	\includegraphics*[width=8.5cm]{dynamics_explosive.eps}
	\psfrag{zrp}{ZRP}
	\includegraphics*[width=8.5cm]{dynamics_normal_v2.eps}
	\caption{Top: state of the system $\{m_i\}$ for $v(m)=(m+0.1)^3$ and $L=100,M=400$ at different times: (a) initial condition with randomly distributed particles, (b) the rise of microscopic clusters of particles, separated by empty sites, (c) the macroscopic cluster (condensate) close to the steady state. Middle: positions of five most occupied sites (red squares for the largest cluster) as a function of time in this model. Bottom: the same plot for ZRP condensation for $u(m,n)=1+3/m$ shows completely different dynamics.	\label{dyn_expl}}
\end{figure}

{\it Model definition:}
The model we consider comprises $M$ particles hopping to the right between
sites of a periodic chain of length $L$ as in the ZRP.
Although partial asymmetry may also be considered, we restrict ourselves here to the case
of totally asymmetric hopping: a particle hops from
site $i$ to site $i+1$ with rate $u(m_i,m_{i+1})$ 
where $m_i, m_{i+1}$  are the occupancies of the 
departure and arrival sites, respectively. We assume the factorized form 
\bq 
	u(m,n) = (v(m)-v(0))v(n), \label{ufact} 
\eq 
where the function $v(m)$ grows as a power of $m$ 
\bq 
	v(m) =(\epsilon+m)^\gamma \sim m^\gamma, \label{cc1}
\eq 
with $\epsilon\ll 1$ and $\gamma>0$ \footnote{The linear case $\gamma =1$
has been studied as an `inclusion process' in Ref.~\cite{gross}.}.  Equation (\ref{ufact}) implies that $u(0,n)=0$
and that for large $m,n$, $u(m,n) \sim (mn)^\gamma$ is the bigger the more particles
are located on {\em both} sites. This has dramatic consequences for
the dynamics.
Comparing simulation results of this model to ZRP dynamics in
Fig.~\ref{dyn_expl} reveals some striking differences (see also
animations in Supp. Material \cite{SUPP}).  In ZRP,  initial microscopic
clusters are first formed, but they coalesce and grow quickly, until  two
macroscopic clusters are left. These slowly merge into the final macroscopic condensate by
exchanging particles through the other sites which form the fluid background. In our model,
particles also aggregate into clusters (see Fig.~\ref{dyn_expl}b) but then these clusters start
to move in the direction of hopping particles. This process speeds up
in time; some clusters move faster as they gain particles in
collisions, and one of them - the condensate - starts to dominate (Fig.~\ref{dyn_expl}c). Due
to the rapid nature of this process we call it explosive
condensation. The speed $v\equiv di_{\rm max}/dt$ at which the condensate travels through the
system stabilizes after the system reaches the steady state. The
motion of the condensate is similar to the ``slinky''-like motion of a
non-Markovian model~\cite{HMS09}. Finally, smaller clusters
move in the opposite direction to the main condensate 
at each collision.

The dynamics thus differs significantly from the zero-range
process. Surprisingly, both models share similar static properties.
In fact, they belong to a class of processes that have the important
property that the steady state probability $P(\{m_i\})$ of a
configuration with $m_1,\dots,m_L$ particles at sites $i,\ldots ,L$ factorizes:
\bq 
	P(\{m_i\})= \prod_{i=1}^L f(m_i), \label{fullfact} 
\eq
with $f(n)$ defined as 
\bq 
	f(n) = f(0) \left(\frac{f(1)}{f(0)}\right)^n \prod_{k=1}^n \frac{u(1,k-1)}{u(k,0)} \label{fn}. 
\eq 
Equation (\ref{fullfact}) requires two conditions on $u(m,n)$ \cite{SUPP}, which are satisfied  for  our model (\ref{ufact}) and  the ZRP (where  $u(m,n)=1+\gamma/m$ for $m>0$ and $u(0,n)=0$). In both cases we can choose $f(1)=f(0)$ and obtain from Eq.~(\ref{fn}) the large $m$ behaviour $f(m)\sim m^{-\gamma}$.
It is known \cite{EH05}  that for a power-law $f(m)$, condensation happens when the density of particles $\rho=M/L$ exceeds the critical density $\rho_c = \lim_{z\to 1} zF'(z)/F(z)$, where $F(z)=\sum_m f(m) z^m$ and $z$ plays the role of fugacity. For $\gamma>2$, $\rho_c$ is finite but for $\gamma<2$, $\rho_c\to\infty$. Therefore, condensation is possible only for $\gamma>2$ and for $\rho>\rho_c$, which marks the transition between condensation/no condensation regimes
\cite{EH05}.

We now come back to the dynamics of our process and investigate
what determines the speed of clusters and the condensate,
how the clusters collide, how long it takes to reach the steady
state, and how this time depends on the initial condition. 
We are interested in the limit of large $M,L$ and fixed density $\rho=M/L$.
For our choice $v(m)=(\epsilon+m)^\gamma$ and $\gamma>2$, we obtain $f(m)\cong
f(0)\epsilon^\gamma m^{-\gamma}$ for $m>0$ and the critical density,
$\rho_c = \sum_m m f(m) /\sum_m f(m)\approx \epsilon^\gamma
\zeta(\gamma-1)\ll 1$ where $\zeta(\gamma-1)$ is the Riemann zeta
function.  As the critical density is low one can make the simplifying
approximation that the clusters move in an otherwise empty system.

Let us first calculate the speed at which the cluster of $m$
particles moves through the system. 
We assume that at $t=0$ the
cluster occupies site $i$, so that $m_i=m$, and that there are no
particles at sites $i-1,i+1$. 
The time $\tau$ it takes to move the cluster
 to site $i+1$ is the sum of times
$t_m,t_{m-1},\dots,t_1$ it takes to move one particle to the right if
the cluster has $m,m-1,\dots, 1$ particles, respectively. Each $n$th hop
is a random process with average duration $t_n$ given by the
inverse of the hopping rate $u(n,m-n)$, thus 
\bq
	\left<\tau\right> = \sum_{n=1}^m \frac{1}{u(n,m-n)}.
\eq 
Recalling that for condensation we are interested in  $\gamma>2$ and using Eqs.~(\ref{ufact}-\ref{cc1}), we obtain that $\left<\tau\right>\approx (\epsilon m)^{-\gamma}$,
which  shows that larger clusters move faster. The condensate, which has $\approx M$ particles, moves $1/\left<\tau\right> \approx (\epsilon M)^\gamma = (\epsilon\rho)^\gamma L^\gamma$ sites per unit time, in agreement with simulations: for parameters from Fig.~\ref{dyn_expl} we have measured the speed $63850\pm 100$ whereas the formula $(\epsilon M)^\gamma$ gives $64000$.

\begin{figure}%
\psfrag{i}{$i$}\psfrag{i+1}{$i+2$}
\psfrag{mi}{$m_i$}\psfrag{mi1}{$m_{i+2}$}
\psfrag{mip}{$m_i'$}\psfrag{mi1p}{$m_{i+2}'$}
\psfrag{yy}{$P(\Delta m)$} \psfrag{xx}{$\Delta m$}
\psfrag{tt}{$t$} \psfrag{mm}{$m$}
\psfrag{(a)}{(a)}
\includegraphics*[width=8cm]{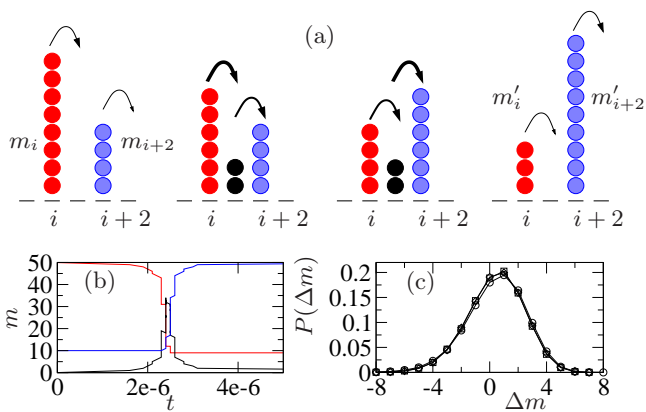}\\
\vspace{3mm}
\psfrag{(b)}{(b)}\psfrag{(c)}{(c)}
\includegraphics*[width=8cm]{collision_distr.eps}
\caption{Top (a): A collision of two condensates having initially $m_i$ and $m_{i+2}$ particles (left) proceeds through exchange of particles at site $i+1$ (middle pictures). Arrows of different sizes show relative magnitudes of hopping rates. After the collision (right), the masses are $m_i'$ and $m_{i+2}'$, with $\left<m_i'-m_i\right><0,\left<m_{i+2}'-m_{i+2}\right>>0$. Bottom left (b): example of stochastic evolution of $m_i(t)$ (red), $m_{i+1}(t)$ (black), and $m_{i+2}(t)$ (blue). Bottom right (c): probability distribution $P(\Delta m)$ of the difference $\Delta m = m_i - m_i'$ for $m_i=25,50,100$ (circles, squares, diamonds) and $m_{i+2}=10,\epsilon=0.1,\gamma=3$. In all cases $\left<\Delta m\right>\approx 0.4$. }%
\label{collision}%
\end{figure}

To understand what happens when two condensates collide with each
other, we assume that a bigger condensate with $m_i$ particles
approaches a smaller one with $m_{i+2}$ particles from the left, and
that they are separated by an empty site $i+1$, see
Fig.~\ref{collision}a. Initially, the dynamics is dominated by hops
from site $i$ to site $i+1$ because $u(m_i,m_{i+1})\propto m_i^\gamma$
is bigger than $u(m_{i+1},m_{i+2})\propto m_{i+2}^\gamma$. As
particles accumulate at site $i+1$, $u(m_{i+1},m_{i+2})$ grows and
$u(m_i,m_{i+1})$ decreases until they become comparable. This happens
when $m_i\approx m_{i+2}$ since $u(m,n)\sim (mn)^\gamma$ is symmetric
for large $m,n$. Then the second half of the process becomes a time-reversed 
and space-inverted version of
itself (see Fig.~\ref{collision}b); one can think of the flow from $i$ to $i+2$ as a flow from $i+2$ to $i$ in reversed time
which is identical to the flow from $i$ to $i+2$ in the first half of the process.
Due to this time-reversal
symmetry, the final configuration becomes the initial configuration
reflected around site $i+1$, modulo random fluctuations. 
In the case of (\ref{cc1}),  the slightly
broken symmetry of $u(m,n)$ produces a small net current of particles
from smaller to bigger clusters (see Fig.~\ref{collision}c).

%Although in this paper we focus on the case exhibiting condensation, $\gamma %>2$, the above analysis, in the particular the dynamics of collisions and the %short-time movement of clusters,
%remians valid also  for $\gamma<2$ for which there is no condensation in our %model for fixed $\epsilon>0$. However, for $\gamma\leq 1$, the velocity of %the cluster $\left<\tau\right>^{-1}\approx m^{2\gamma-1}$ does not depend on %$\epsilon$. Therefore, we expect the same dynamics also in the inclusion %process \cite{gross} which is a special case of our model with $\gamma=1$. 

We now venture to draw the following picture of condensation
dynamics. First, small clusters are formed randomly from the initial
state. For a system of size $L$ there will be $N=O(L)$ such
clusters. Subsequently, these clusters move ballistically between collisions,
which are almost elastic. One of them soon collects more particles
than the rest and starts moving at increasing speed, gaining mass and
becoming the final condensate. Let us calculate the time $T_{\rm ss}$
for the system to relax to stationary state. Each cluster will go
through a series of collisions and either dissolve into the background
or become the condensate; in either case we can associate a relaxation
time $T$ to each cluster (with $T=\infty$ if the cluster disappears).
Then $T_{\rm ss}$ will be the minimal time out of $T_1,\dots,T_N$ relaxation times for
all $N$ clusters: \bq T_{\rm ss} = \min \left\{ T_1,\dots, T_N
\right\}. \label{tss} \eq The relaxation process of a particular
cluster of initial mass $m_0$ is a series of  transitions at times $t_n$ at which it moves by one site to the right and (possibly) exchanges a chunk of mass $\Delta m_n$ with other clusters:
\ba
	m_n &=& m_{n-1} + \Delta m_n, \\
	t_n &=& t_{n-1} + \Delta t_n.
\ea
Here $\Delta t_n$ is the time between two jumps and is exponentially distributed as
\begin{equation}
	p_n(\Delta t_n) = \lambda_n e^{-\lambda_n \Delta t_n}, \label{pndt}
\end{equation}
where $\lambda_n\propto m_n^\gamma$ is the speed of the cluster.
Let us calculate the probability distribution $f(T)$ of the relaxation time $T=\Delta t_1+ \Delta t_2+\dots$. 
Numerical simulations suggest that the mass $m_n$ increases linearly through the collisions. 
We may thus assume that $m_n \propto  n$ and $\lambda_n \simeq  A n^{\gamma}$ for large $n$.
\begin{figure}
	\centering
	\psfrag{xx1}{$T$} \psfrag{yy1}{$f(T)$}
	\psfrag{xx}{$T_{\rm ss}$} \psfrag{yy}{$h(T_{\rm ss})$}
		\includegraphics*[width=8cm]{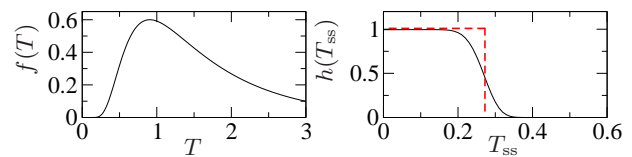}
	\caption{\label{fig:exft+stepfunc}Sketches of $f(T)$ (left) and $h(T_{\rm ss})=\exp(-N\int_0^{T_{\rm ss}} f(T) dT)$ (right). $h(T_{\rm ss})$ can be approximated by a step function (see text). }
\end{figure}
Then $T$ is a sum of independent exponential random variables
and $f(T)$ is given by
\begin{equation}
	f(T) = \frac{1}{2\pi} \int_{-\infty}^{\infty} d\omega e^{-i\omega T} \tilde{f}(\omega),
\label{fT}
\end{equation}
where $\tilde{f}(\omega)$ is the product of characteristic functions of exponential distributions (\ref{pndt}):
\begin{equation}
	\tilde{f}(\omega) = \prod_{n=1}^\infty 
\tilde{p}_n(\omega) = \prod_{n=1}^\infty  \frac{1}{1-i\omega/\lambda_n}. \label{fomega}
\end{equation}
We expect that $f(T)$ has the shape depicted in Fig.~\ref{fig:exft+stepfunc} and that it decays to zero for $T\to 0$.
The large-$\omega$ behaviour of $\tilde{f}(\omega)$, which corresponds to small-$T$ behaviour of $f(T)$, is given by \cite{SUPP}
\begin{equation}
	\tilde{f}(\omega) \cong - i (2\pi)^{\gamma/2} \sqrt{\frac{i\omega}{A}} \exp\left[ - \frac{2\pi i (i\omega/A)^{1/\gamma}}{e^{2\pi i/\gamma}-1}  \right].
\end{equation}
Now, we must  invert the Fourier transform to recover $f(T)$. 
For small $T$, 
this may be done by the saddle point approximation 
to the integral over $\omega$ (dominated by $\omega =O( T^{-\gamma/(\gamma-1)})$)
and one obtains
\begin{equation}
	f(T)  \propto C T^{(1-3\gamma)/(2(\gamma-1))} \exp\left[-B (AT)^{-\frac{1}{\gamma-1}}\right], \label{f(T)}
\end{equation}
where $B,C$ are some real, positive constants.
If we assume that each cluster evolves independently,
the relaxation time (\ref{tss}) of the system becomes the minimum  out of $N$ 
independent random variables distributed according to $f(T)$.
Extreme values statistics tells us that the distribution $P(T_{\rm ss})$
is given by 
\bq
	P(T_{\rm ss}) = Nf(T_{\rm ss}) \left[ \int_{T_{\rm ss}}^\infty f(T) dT \right]^{N-1}, 
	\label{Pss2}
\eq
and integrating by parts and expanding for $f(T)$ small,
\bq
	\left< T_{\rm ss}\right> 
	\cong \int_0^\infty \exp\left(-N\int_0^{T_{\rm ss}} f(T) dT\right) d T_{\rm ss} . \label{avertts}
\eq
Knowing the small-$T$ behaviour  (\ref{f(T)}) of $f(T)$, we can calculate the average (\ref{avertts}) 
for large $N$ as follows.
The function $h(T_{\rm ss})=\exp(-N\int_0^{T_{\rm ss}} f(T) dT)$ approaches a step function for large $N$, see Fig.~\ref{fig:exft+stepfunc}. The integral (\ref{avertts}) over $T_{\rm ss}$,
then becomes 
$
\left<T_{\rm ss}\right>  \cong
\int_0^\infty h(T_{\rm ss}) d T_{\rm ss} \cong t_0,
$
where $t_0$
is the position of the step in $h$, 
which can be  identified as  the point
at which  $h''=0$, yielding
\bq
f'(t_0) = Nf^2(t_0)\;.
\eq
Inserting the short-time behaviour  (\ref{f(T)}) of $f$ into 
this condition one obtains
\begin{equation}
	CN(\gamma-1) = B\exp\left[B (At_0)^{-\frac{1}{\gamma-1}}\right] A^{-\frac{1}{\gamma-1}} t_0^{-\frac{\gamma}{\gamma-1}}.
\end{equation}
Taking logarithms yields
\begin{equation}
t_0 \simeq \left( \frac{1}{\beta}\left[
\ln N - \ln \left( \frac{\beta}{\gamma -1}\right) - \frac{1}{2} \ln t_0
\right]\right)^{1-\gamma}.
\end{equation}
Thus, recalling $N= O(L)$ and 
$\left<T_{\rm ss}\right>  \cong t_0$, the relaxation time asymptotically decreases as
\begin{equation}
	 \left<T_{\rm ss}\right> = c_2 (c_3 + \ln L)^{1-\gamma}. \label{Tfit}
\end{equation}
This form crosses over from 
$\left<T_{\rm ss}\right> = 1/({\rm const}  + O(\ln L))$
for small $L$ to
$\left<T_{\rm ss}\right> = O( (\ln L)^{1-\gamma} )$
for large $L$. This differs much from ZRP-like models where $\left<T_{\rm ss}\right> \sim L^2$ grows with $L$ \cite{EH05,godreche+luck}.
Since the time to steady state decreases with $L$, an infinite system relaxes instantaneously. This is reminiscent of instantaneous gelation known from the theory of coagulation processes \cite{dongen}. 
In fact,  our model provides a non-trivial example of  instantaneous gelation
in a spatially-extended system.
However, the model
and its effective description in terms of colliding clusters
 differ from coagulation processes in that there is exchange of particles between 
clusters rather than coagulation (a model with exchange of particles has been studied in Ref.~\cite{ben-naim-krapivsky}, see Supp. Material for more details). 

\begin{figure}
\psfrag{xx}{$L$} \psfrag{yy}{$\left<T_{\rm ss}\right>^{-1}$}
	\includegraphics[width=8.5cm]{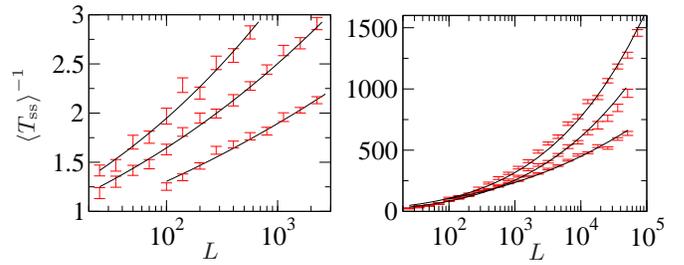}
	\caption{\label{th_vs_exp}$\left<T_{\rm ss}\right>^{-1}$ obtained in numerical simulations (points) and from formula (\ref{Tfit}) fitted to data points (lines). In all cases the density $\rho=2$ and $\gamma=3,4,5$ (curves from bottom to top). Left: $v(m)=(0.3+m)^\gamma$, every 5th site has initially 10 particles. Right: $v(m)=(1+m)^\gamma$ particles are distributed randomly in the initial state. $\left<T_{\rm ss}\right>^{-1}$ for different $\gamma$ differ by orders of magnitude and hence they have been rescaled to plot them together.}
\end{figure}

In the above derivation we assumed that $\lambda_n$ is strictly proportional to $n^\gamma$, and that the proportionality coefficient is the same for all clusters. This is valid only if all clusters have the same initial size $m=1$. To account for fluctuations of cluster sizes one should take the product (\ref{fomega}) not from $n=1$ but from some $n_0>0$, 
%\begin{equation}
%	\tilde{f}(\omega) \approx \left[\prod_{n=n_0}^\infty \left(1-\frac{i\omega}{An^\gamma}\right)\right]^{-1}, %\label{infproduct}
%end{equation}
with $n_0$ changing from cluster to cluster. However, this does not modify the asymptotic behaviour of
$\tilde{f}(\omega)$, it only increases the constant $c_3$ in Eq.~(\ref{Tfit}). We have checked numerically evaluating Eqs.~(\ref{fomega}), (\ref{fT}) and (\ref{avertts}) that $\left<T_{\rm ss}\right>$ for $n_0>1$ has much stronger finite-size corrections and behaves as $\sim 1/({\rm const}  + O(\ln L))$ for a wide range of $L$. Although $c_3$ may  in principle be calculated from our theory for $n_0>1$, in practice it is simplest to treat  $c_3$  as a free parameter. In this way  Eq.~(\ref{Tfit}) fits numerical simulations very well.  To check this, we measured the time it took the biggest cluster to reach the mean steady-state size of the condensate, $M-L\rho_c$. In Fig.~\ref{th_vs_exp} we compare Eq.~(\ref{Tfit}) with $\left<T_{\rm ss}\right>$ obtained in simulations, for different initial conditions. We plot $\left<T_{\rm ss}\right>^{-1}$ because it shows convincingly that $\left<T_{\rm ss}\right>^{-1}$ grows to infinity for $L\to\infty$, and therefore $\left<T_{\rm ss}\right>\to 0$ in this limit.

In conclusion, we have elucidated a form of dynamic condensation which happens in far-from-equilibrium system of hopping particles. % with factorized steady states. 
In contrast to previously studied models, the condensate moves through the system and 
its dynamics  speeds up in time---hence we term the condensation ``explosive''. The relaxation  is dominated by the  process of initial coalescence which is the slowest stage of condensate formation, at variance with 
previously studied models of condensation such as the ZRP where this stage is the fastest. It remains to be seen whether condensation can be made ``explosive'' also in models which do not have a factorized steady state, such as those with spatially extended condensates \cite{evans1}.

\end{document}